\documentclass[aip,pra,reprint,amsfonts,amsmath,amssymb,superscriptaddress,10pt,floatfix]{revtex4-1}

\usepackage[english]{babel}
\usepackage[utf8]{inputenc}

\usepackage{amsmath}
\usepackage{amsfonts}
\usepackage{amssymb}
\usepackage{amsthm}
\usepackage{mathtools}
\usepackage{graphicx}
\usepackage{xfrac}
\usepackage{nicefrac}

\usepackage{color}
\usepackage{array}
\usepackage{longtable}
\usepackage{calc}
\usepackage{multirow}
\usepackage{hhline}
\usepackage{ifthen}

\usepackage{times}
\usepackage{mathptmx}

\usepackage[colorlinks]{hyperref}

\graphicspath{{figures/}}

\hypersetup{citecolor=blue, urlcolor=blue, linkcolor=blue, colorlinks=true}

\begin{document}

\title{Nucleation and structural growth of cluster crystals}

\author{Christian Leitold}
\author{Christoph Dellago}

\affiliation{Faculty of Physics, University of Vienna, Boltzmanngasse 5, 1090 Vienna, Austria}

\date{\today}

\begin{abstract}
We study the nucleation of crystalline cluster phases in the generalized exponential model with exponent $n=4$. Due to the finite value of this pair potential for zero separation, at high densities the system forms cluster crystals with multiply occupied lattice sites. Here, we investigate the microscopic mechanisms that lead to the formation of cluster crystals from a supercooled liquid in the low-temperature region of the phase diagram. Using molecular dynamics and umbrella sampling, we calculate the free energy as a function of the size of the largest crystalline nucleus in the system, and compare our results with predictions from classical nucleation theory. Employing bond-order parameters based on a Voronoi tessellation to distinguish different crystal structures, we analyze the average composition of crystalline nuclei. We find that even for conditions where a multiply-occupied fcc crystal is the thermodynamically stable phase, the nucleation into bcc cluster crystals is strongly preferred. Furthermore, we study the particle mobility in the supercooled liquid and in the cluster crystal. In the cluster crystal, the motion of individual particles is captured by a simple reaction-diffusion model introduced previously to model the kinetics of hydrogen bonds.
\end{abstract}

\maketitle

\section{Introduction}

Everyday life tells us that in the macroscopic world two objects cannot occupy the same part of space. Similarly, on the microscopic level, Fermi repulsion prevents atoms from sharing the same location. However, exactly that can happen on the mesoscopic level. Macromolecules such as polymer chains, polymer rings, or dendrimers, can penetrate each other in such a way that their respective center of mass positions coincide~\cite{Sciortino2013, Mladek2008, Lenz2011, Lenz2012}. In contrast to atoms and compact molecules, the effective interaction of two such open molecules will remain finite even for zero separation making it possible for the molecules to overlap. This feature is mimicked by coarse grained models in which every macromolecule is represented by a single particle~\cite{Likos2008, Mladek2008a, Zhang2010, Coslovich2011, Coslovich2012, Nikoubashman2012, Wilding2014}. One example for such a potential is the \emph{generalized exponential model} or GEM-$n$ system, where the pair interaction is given by an exponential,
\begin{equation}
u(r) = \varepsilon e^{- (r / \sigma)^n}.
\end{equation}
This potential permits the existence of cluster phases for $n~>~2$, indicated by the presence of negative components in the potential's Fourier transform~\cite{Likos2001}. In particular, for the case $n=4$ and $d=3$ dimensions, there exists a very complex phase behavior showing a multitude of cluster phases in the low-temperature regime~\cite{Zhang2010, Zhang2012, Wilding2014}. For low densities, the system is in a liquid state. When increasing the density, for very low temperatures, the system initially forms a regular face-centered cubic (fcc) or body-centered cubic (bcc) lattice, depending on the exact conditions. A further increase in the density leads to the formation of cluster crystals, with the number of particles per lattice site increasing with the density. The properties of clusters phases in GEM-4 have also been studied extensively in $d=2$ dimensions~\cite{Montes-Saralegui2015, Prestipino2014,  Archer2014}.

In this paper, we investigate the formation of cluster crystals through nucleation out of the metastable liquid in the GEM-$4$ system, finding the crystallization occurs predominantly into the bcc phase. Similar preference for bcc was observed recently in the Gaussian core model (GEM-2), which, however, does not form clusters~\cite{Lechner2011, Mithen2015}.

In GEM-4 as well as in the Gaussian core model, crystallization results in a high degree of polymorphism. On the other hand, a rather different nucleation mechanism has been observed in a binary Lennard-Jones mixture~\cite{Jungblut2011}. There, in accordance with Ostwald's “step rule”~\cite{Ostwald1897}, a core primarily consisting of fcc particles (the stable phase) is surrounded by a surface layer of bcc particles (the phase with the lowest free energy difference to the liquid).

At this point, an explanation of the terminology used in this paper is in order. We use the word \emph{particle} when speaking about individual, soft particles, and \emph{cluster} when speaking about a small aggregation or “blob” of individual particles \emph{sitting on top of each other}.  With \emph{cluster occupancy} we mean the number of individual particles per cluster. The word \emph{nucleus}, on the other hand, is reserved for larger aggregations of either particles or clusters, arranged in a regular manner on a lattice and forming the precursor of an extended crystal.

The remainder of this paper is structured as follows. In Sec.\,\ref{sec:model_gem4}, we will describe the model and briefly discus the regions of the phase diagram relevant for this work. Sec.\,\ref{sec:methods_gem4} covers the numerical methods used in this study, such as the algorithm to assign individual soft particles to clusters and the details of the dynamics used in our work. In the same section, we explain our method for detecting crystal structures using bond-order parameters based on a Voronoi tessellation of space. We present our results in Sec.\,\ref{sec:results_gem4} and provide a discussion in Sec.\,\ref{sec:discussion_gem4}.

\section{Model}
\label{sec:model_gem4}

\begin{figure}
\includegraphics[width=0.6\columnwidth]{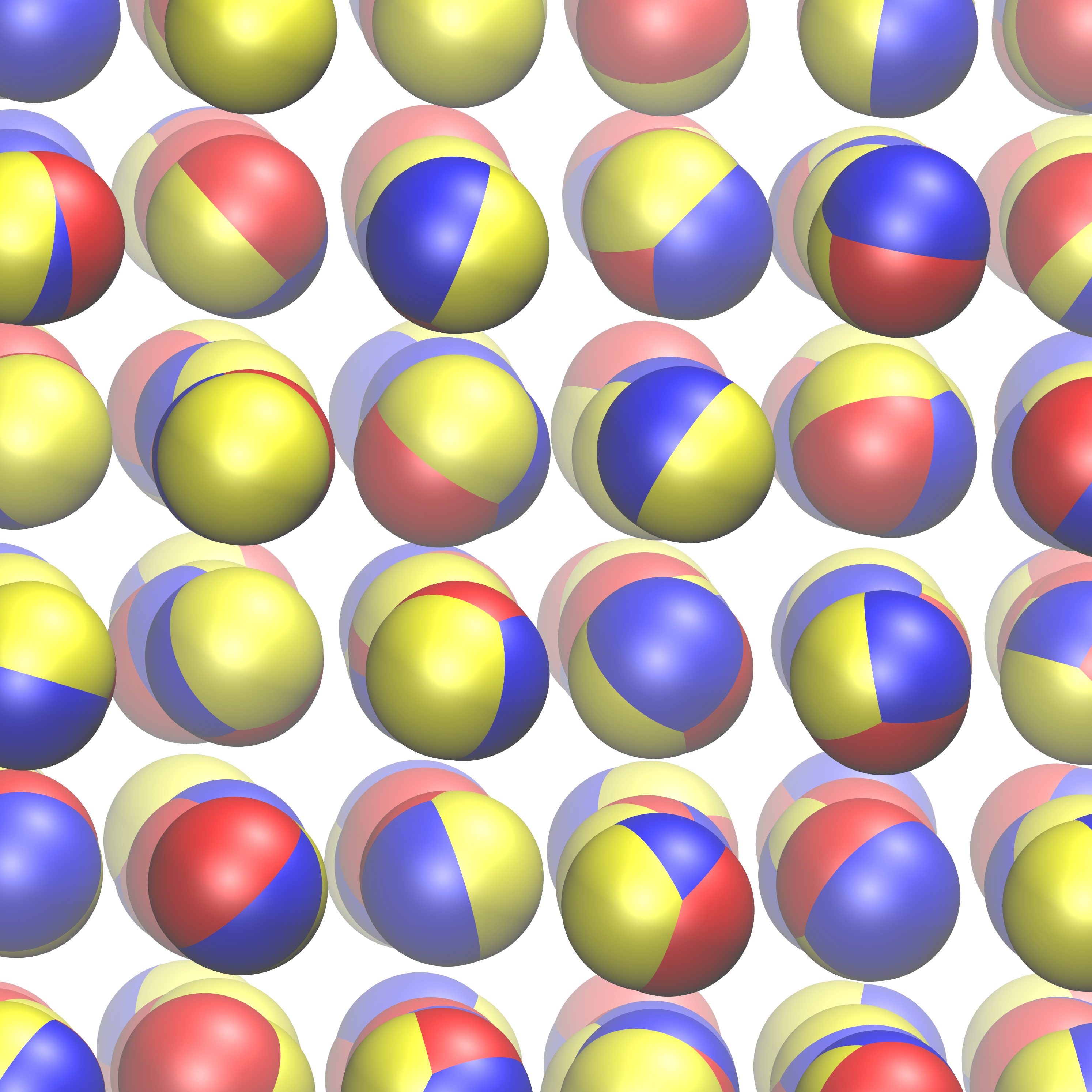}
\caption{GEM-4 cluster crystal in an fcc3 structure at low temperature. Each lattice site of the face-centered cubic crystal structure is occupied by three particles. The particles are displayed with three different colors in order to enhance the visibility of the individual cluster members.}
\label{fig:plot_pov}
\end{figure}

We study the cluster-forming GEM-4 system of spherical particles interacting via the pairwise  additive potential
\begin{equation}
U(\mathbf{r}_1, \dots, \mathbf{r}_N) = \sum_{i<j} \varepsilon e^{- (r_{ij} / \sigma)^4},
\end{equation}
where $r_{ij} = |\mathbf{r}_i -\mathbf{r}_j|$ is the distance between particles $i$ and $j$, and $\varepsilon$ and $\sigma$ set the energy and length scale, respectively. This system shows the formation of cluster crystal phases at high densities~\cite{Zhang2010}. For very low temperatures, the system is in a face-centered cubic structure where the number of individual particles per lattice site increases with the density. Provided that the dimensions of the simulation box are chosen commensurate with an fcc cluster crystal, at these conditions, there is very little variance in the occupation numbers between different sites. Thus, each site is occupied by $m$ particles, where $m$ is called cluster occupancy. We denote this structure by fcc$m$. As an example, consider the fcc3 crystal at a rather low temperature of $k_{\text{B}} T / \varepsilon = 0.02$ shown in Fig.\,\ref{fig:plot_pov}, at which the thermal fluctuations of the particle positions are very small. At higher temperatures, the system is in a cluster-fcc or cluster-bcc structure. Under these conditions, the mean cluster occupancy also increases with density, but is in general a non-integer number, i.\,e., the number of individual particles typically differs from one lattice site to another. Here, one also observes frequent hops of individual particles between clusters. The system forms clusters also in the liquid phase at sufficiently high densities and it is from such a cluster liquid that we observe crystal nucleation. We restrict ourselves to situations where the mean cluster occupancy is an integer, typically two, and does not vary significantly between clusters. For the comparatively low temperatures investigated in this work, this is true despite the fact that strictly speaking, we are in the cluster-fcc and cluster-bcc regions of the phase diagram, respectively.

\section{Methods}
\label{sec:methods_gem4}

In this section, we describe the algorithm used to group single particles into clusters and then how we use bond-order parameters based on a Voronoi tessellation of space to distinguish between liquid and different crystal environments. We also give a detailed description of the order parameter used in our work, the size of the largest crystalline nucleus. The section is concluded by a discussion of the dynamics employed in our simulations.

\subsection{Clustering algorithm}
\label{sec:clustering}

\begin{figure}
\includegraphics[width=1.0\columnwidth]{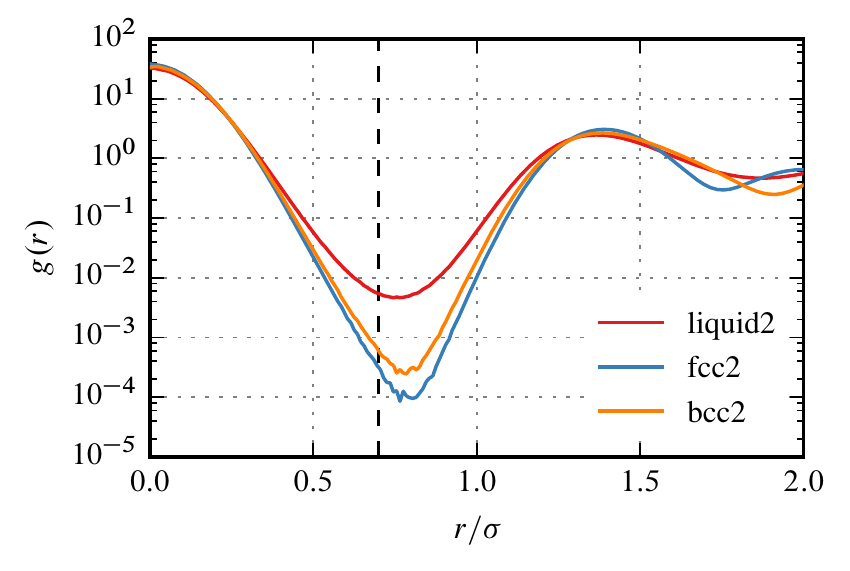}
\caption{Radial distribution function $g(r)$ for different phases at $k_{\text{B}} T / \varepsilon = 0.08$ and $\rho \sigma^3 = 1$ (stable phase: fcc2). The cutoff radius $r_c = 0.7 \sigma$ used to define clusters is indicated by a  vertical dashed line. Note the logarithmic scale on the $y$-axis.}
\label{fig:gr_inner}
\end{figure}

Consider the typical form of the radial distribution function $g(r)$ for the GEM-4 system as shown in Fig.\,\ref{fig:gr_inner}. The high peak for zero separation indicates that clusters form at high density. Thus, in order to define what constitutes a cluster of particles, we first find, for each particle, all neighboring particles within a distance of $r_c = 0.7 \sigma$. Then, we find all clusters of particles connected by that distance criterion~\cite{Stillinger1963}. It is worth noting that by using this method, it is theoretically possible that within a single cluster, two particles are separated by a distance larger than $r_c$. As a simple example, one might think of three particles arranged along a line, such that the distance between the first and the third particle is significantly larger than the cutoff radius. However, in practice, in GEM-4 under the conditions investigated in this study, this rarely happens: typical cluster occupancies are on the order of two, and even clusters of three particle are rather closely packed.

After the clustering step, we calculate the center-of-mass position for each cluster. In order to avoid any ambiguities arising from the periodic boundary conditions employed in our simulation, we do this by mapping each coordinate onto a unit circle. A detailed description of the algorithm is given in Appendix\,\ref{ap:com_pbc}. We use this cluster center-of-mass coordinates exclusively in the following procedures to search for neighbors, calculate bond-order parameters, and detect crystal structures. Trivially, for single-particle “clusters”, we just use the raw coordinates.

\subsection{Voronoi-based bond-order parameters}
\label{sec:voronoi}

Since we want to study crystal nucleation within a predominantly liquid system, we need a way to distinguish liquid from solid particles. Furthermore, we would also like to find out whether any single solid particle is say in an fcc, bcc, or hcp environment. For that task, we use a variation of the bond-order parameters suggested in Ref.\,\onlinecite{localbondorder}.

The usual expression for the complex bond-order vector of particle (or cluster) $i$ is~\cite{steinhardt}
\begin{equation}
q_{lm}(i) = \frac{1}{N_b(i)} \sum_{j} Y_{lm}(\mathbf{r}_i - \mathbf{r}_j),
\end{equation}
where the $Y_{lm}$ are the spherical harmonics and the sum runs over all neighbors, or \emph{bonds}, of particle $i$. In the expression above it is not specified how neighbors are actually defined. A traditional choice is to view all particles within some cutoff radius around particle $i$ as neighbors. However, more recently, a fixed number of neighbors has been used, e.\,g., the closest 12 neighbors, as this adapts better to varying local densities~\cite{Russo2012, Mithen2015}. Similarly, solid-angle based nearest neighbors~\cite{VanMeel2012} also cope well with varying local environments. An alternative approach, applied in this work, is to use a Voronoi tessellation of space in order to define neighbor relations~\cite{Mickel2013}. One defines two particles as neighbors if their corresponding Voronoi cells share a common facet. This has several advantages. First of all, the method is completely parameter-free, thus eliminating the need to fine tune parameters, as it is typically the case when employing a cutoff-based neighbor definition. Also, similarly to the method using a fixed number of neighbors and the solid-angle based method, the Voronoi-based method works for all densities. In addition, one can weight the contribution of each neighboring particle to the complex bond-order vector by its corresponding Voronoi facet area in order to make the order parameter a continuous function of the particle coordinates. This new Voronoi-based complex bond-order vector can be written as
\begin{equation}
q_{lm}^{v}(i) = \sum_{f \in \mathcal{F}(i) } \frac{A(f)}{A} Y_{lm}(\Theta_f, \varphi_f).
\end{equation}
Here, the sum runs over all facets $f$ of the Voronoi cell $\mathcal{F}$ associated with particle $i$, $A(f)$ is the corresponding facet area, and $A$ is the total facet area of the cell. The unit normal vector of each facet, defining the angles $\Theta_f$ and $\varphi_f$ on a unit sphere, coincides with the (normalized) bond vector for the corresponding neighbor. Therefore, from an algorithmic point of view, this Voronoi-based calculation of bond-order parameters is very similar to the usual one. The only two differences are that one uses the Voronoi construction to find neighbors, and weights each neighbor contribution by the respective facet area. In our simulation, we used the open-source library Voro++~\cite{Rycroft2009}, version 0.4.6, to perform the calculation of the Voronoi cells, neighbor relations, and facet areas.

Next, we calculate a modified version of the averaged bond order parameters of Ref.\,\onlinecite{localbondorder},
\begin{equation}
\bar{q}_{lm}^{v}(i) = \frac{\frac{A}{12}  q_{lm}^{v}(i) + \sum_{f \in \mathcal{F}(i) } A(f) q_{lm}^{v}(f)}{\frac{A}{12} + A} .
\end{equation}
That is, instead of weighting the central particle and each neighboring particle equally, we again make use of the corresponding Voronoi facet areas as weights for the neighbor contributions. Note that the central particle is weighted by $\nicefrac{1}{12}$ of the total surface area. This ensures that the values are comparable to the traditional definition of averaged bond-order parameters, since in the case that neighbors are defined by a cutoff distance, often the mean number of neighbors is close to 12 as well.

Finally, we calculate the local bond-order parameters
\begin{equation}
\bar{q}_l^{v}(i) = \sqrt{ \frac{4 \pi}{2 l + 1} \sum_{m=-l}^{l} |\bar{q}_{lm}^{v} (i)|^2 }
\label{eq:qi}
\end{equation}
and
\begin{equation}
\bar{w}_l^{v}(i) = 
\frac{\displaystyle \sum \limits_{m_1 + m_2 + m_3 = 0} 
\begin{pmatrix}
      l & l &  l \\
      m_1 & m_2 & m_3
\end{pmatrix}
\bar{q}_{l m_1}^{v} \! (i) \, \bar{q}_{l m_2}^{v} \! (i) \, \bar{q}_{l m_3}^{v} \! (i)
}
{\left(\displaystyle \sum \limits_{m=-l}^{l} |\bar{q}_{lm}^{v}(i)|^2 \right)^{3/2}}.
\label{eq:wi}
\end{equation}

\begin{figure}
\includegraphics[width=1.0\columnwidth]{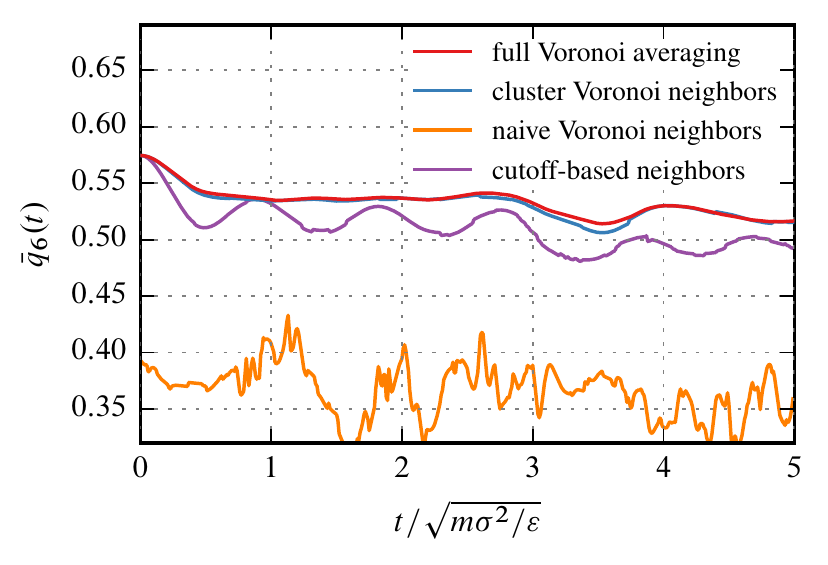}
\caption{Value of $\bar{q}_6$ and $\bar{q}_6^{v}$, respectively, as a function of simulation time for an arbitrary particle in an fcc2 structure using different methods to define the neighbor relations and calculate order parameters. The simulation starts from a perfect fcc2 lattice. The usage of $\bar{q}_6^{v}$ (top line, red), Eq.\,\eqref{eq:qi}, ensures a smoothly varying order parameter without any discontinuous jumps. The traditional, cutoff-based method to define neighbors produces small jumps in the time evolution of $\bar{q}_6$ (purple line). Note also that a naive Voronoi tessellation, applied to raw particle coordinates in a cluster crystal, leads to very unsatisfactory results (bottom line, orange).}
\label{fig:averaging_comparison}
\end{figure}

It is interesting to note that, in principle, the resulting complex vector, and hence the $\bar{q}_l$ and $\bar{w}_l$ numbers derived from it, are continuous in all the particle positions. That is certainly not true when using equal weights and a hard neighbor cutoff, as any bond-order parameter will show a discontinuous jump as soon as a particle enters or leaves the neighbor shell. A similar situation occurs when using a fixed number of neighbors, as the identity of the furthest particle can switch from one simulation time step to another. For cluster-forming particles, of course, a discontinuity is introduced by the cutoff-based clustering procedure described in Sec.\,\ref{sec:clustering}. However, the exact degree of this discontinuity depends on the variance in the cluster occupancy. As an illustration for a low variance in the cluster occupancy, consider Fig.\,\ref{fig:averaging_comparison}, in which we have plotted the time evolution of $\bar{q}_6^{v}$ and $\bar{q}_6$ for an arbitrary particle in an fcc2 crystal. For the latter case, the effects of three different neighbor definitions are shown: the traditional cutoff-based definition (with a second inner cutoff radius of $0.7 \sigma$ to avoid particles in the same cluster), a naive Voronoi tessellation, applied to the raw particle positions without the cluster averaging procedure described in Sec.\,\ref{sec:clustering}, and an intermediate method, where we use the Voronoi facet areas only in the computation of $q_{lm}$, but not for the neighbor-averaged $\bar{q}_{lm}$. As it turns out, only the fully Voronoi-averaged $\bar{q}_6^{v}$ show a continuous time evolution without any jumps.

One should not forget to mention that the Voronoi-based neighbor detection has one significant disadvantage, namely its higher computational cost. In our experience, calculating order parameters employing the Voronoi tessellation takes roughly five times as long compared to the cutoff-based method. However, in a typical application, this is not a big issue, because the overwhelming majority of the computation time is spent determining forces rather than calculating order parameters. Furthermore, additional properties of the computed Voronoi cells, like the total area and the cell volume, can be useful to analyze the local structure as well.

\subsection{Detection of crystal structures}

\begin{figure}
\includegraphics[width=1.0\columnwidth]{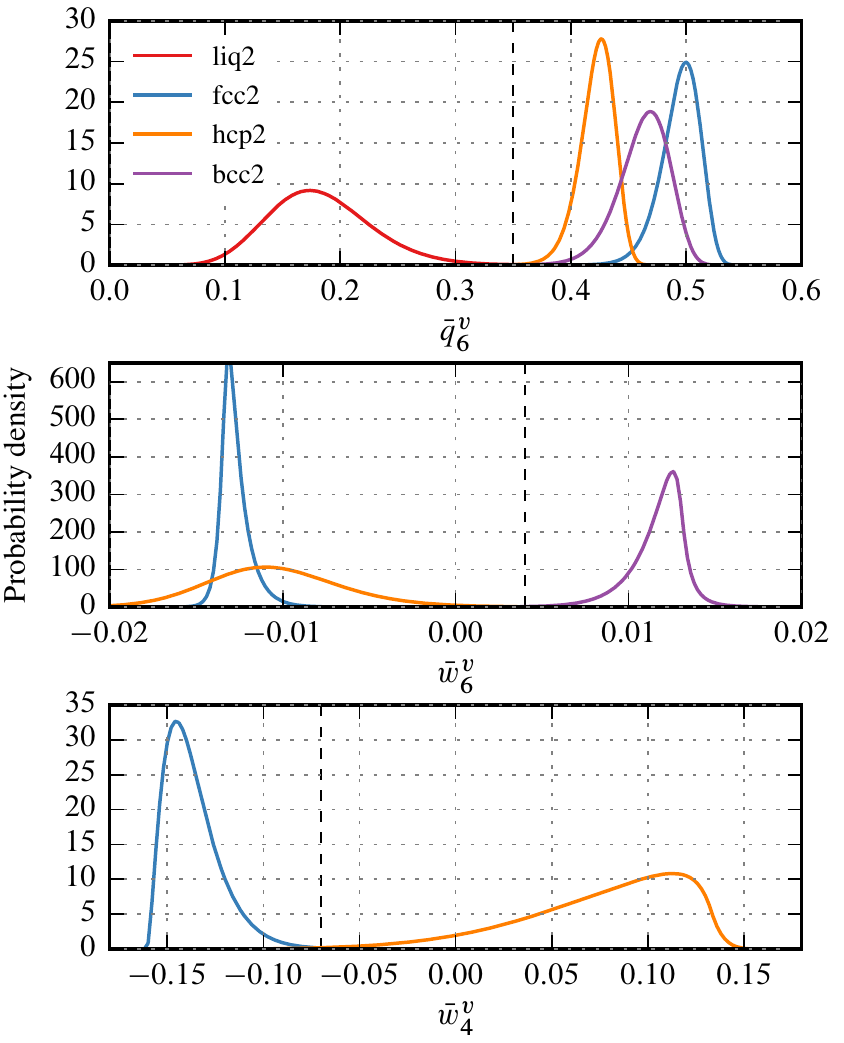}
\caption{Probability densities for $\bar{q}_6^{v}$, $\bar{w}_6^{v}$, and $\bar{w}_4^{v}$ for different pure structures at $k_{\text{B}} T / \varepsilon = 0.08$ and $\rho \sigma^3 = 1$. The decision thresholds described in the text are indicated by vertical dashed lines.}
\label{fig:qw_histos_voronoiaveraging}
\end{figure}

In order to decide whether a particle, or rather, cluster, is in a liquid or crystalline environment, we adopt the method used in recent work of Mithen, Callison, and Sear~\cite{Mithen2015}. It is based on the values of the local bond-order parameters $\bar{q}_6^{v}$, $\bar{w}_6^{v}$, and $\bar{w}_4^{v}$. First, if $\bar{q}_6^{v}(i) < 0.35$, cluster $i$ is classified to be in a liquid-like environment. Otherwise, if $\bar{w}_6^{v}(i) > 0.004$, the cluster is classified as bcc. In the remaining case, we check if $\bar{w}_4^{v} > -0.07$ in order to classify the cluster as fcc or hexagonal close-packed (hcp). This step-wise decision procedure is illustrated in Fig.\,\ref{fig:qw_histos_voronoiaveraging}. While the histograms are shown for a certain temperature and density, they are very similar for all conditions investigated in this work, and hence the same threshold values can be applied for all cases. Also, this set of criteria is very easy to implement and perfectly separates the various structures. In contrast, a previously used similar criterion, based on the $\bar{q}_4$--$\bar{q}_6$ plane~\cite{Lechner2011}, shows a considerable amount of overlap especially between bcc and hcp already for the pure structures in a GEM-4 cluster crystal.

\subsection{Definition of the largest crystalline nucleus}

In order to find the largest crystalline nucleus, we first identify all solid clusters. Then, the largest crystalline nucleus is defined as the largest connected structure of solid clusters.

Any single cluster which is detected as either fcc, hcp, or bcc by the procedure described above is immediately classified as solid. However, for clusters  which are in a liquid environment according to their $\bar{q}_6^{v}$ value, we employ an additional test. In this case, we apply the ten-Wolde-Frenkel criterium\cite{TenWolde1996} by first calculating the connection coefficient between clusters $i$~and~$j$,
\begin{equation}
d_{ij} = \frac{\sum_{m=-6}^6 \; q_{6m}^v(i) \; q^{v*}_{6m}(j)}{ \left( \sum_{m=-6}^6 \; |q_{6m}^v(i)|^2 \right)^{1/2} \; \left( \sum_{m=-6}^6 \; |q_{6m}^v(j)|^2 \right)^{1/2}},
\end{equation}
where $^*$ denotes complex conjugation. The particles are defined as \emph{connected} if $d_{ij} > 0.5$. Then, any cluster with more than $N_f = 8$ connected neighbors is classified as solid. As a result, we are able to overcome a significant disadvantage of a criterion based purely on the values of neighbor-averaged local bond-order parameters. In particular, due to the spatial averaging, a cluster near a liquid--solid interface can have a comparatively low $\bar{q}_6^{v}$ value, even though it is already in a regular arrangement with its neighbors. In contrast, clusters deep within the liquid phase have both a low $\bar{q}_6^{v}$ value and a low number of connections~\cite{Lechner2011}. Using the procedure described above, we can decompose the size of the largest crystalline nucleus $N_{\text{nuc}}$ into a sum of its constituents,
\begin{equation}
N_{\text{nuc}} = N_{\text{liq2}} + N_{\text{fcc2}} + N_{\text{hcp2}} + N_{\text{bcc2}}.
\end{equation}

\subsection{Dynamics and umbrella sampling}

We perform molecular dynamics simulations in a cubic, periodic box for constant particle number $N$ and temperature $T$. Two different algorithms are used to simulate the system both at constant volume (\textit{NVT} ensemble) and at constant pressure (\textit{NpT} ensemble).

For the \textit{NVT} time evolution, we employ Langevin dynamics with a time step $\Delta t = 0.01 \, \sqrt{m \sigma^2/\varepsilon}$ and a relatively low damping constant $\gamma = 0.2 \, \sqrt{\varepsilon / m \sigma^2}$, where $m$ is the particle mass (set to unity in reduced units). Note that the relaxation time is $\tau = 1 / \gamma~=~5 \, \sqrt{m \sigma^2/\varepsilon} = 500 \, \Delta t$. The low damping constant, while still providing sufficient thermostating, ensures that the short-time evolution of the system is hardly perturbed by the friction and random force terms. The actual integration is performed using the Langevin thermostat by Schneider and Stoll~\cite{schneider} implemented in a modified version of LAMMPS~\cite{lammps}, which we call as a library from our custom simulation code.

We perform \textit{NpT} calculations using the same time step. Within LAMMPS, a Nosé--Hoover chain is used for the time integration~\cite{Tuckerman1992}. The relaxation times for temperature and pressure are set to $\tau_T / \sqrt{m \sigma^2/\varepsilon} = 1$ and $\tau_p / \sqrt{m \sigma^2/\varepsilon} = 10$, corresponding to 100 and 1000 time steps, respectively.

We employ umbrella sampling~\cite{TorrieValleauJCP1977} to calculate nucleation free energies as a function of the number of clusters in the largest crystalline nucleus $N_{\text{nuc}}$. 95~windows each spaced a distance of 5~clusters from its neighbors are used. The bias potential for window $j$ is given by
\begin{equation}
U_{\text{bias}}^j(N_{\text{nuc}}) = \frac{k}{2} (N_{\text{nuc}} - 5j)^2,
\end{equation}
where $j = 1, \dots, 95$ and the bias spring constant is set to $k = 0.002\,\varepsilon$. Between evaluations of the umbrella potential, in order to generate a new trial configuration, the system is evolved in time using the $NpT$ scheme described above. Typically, we integrate for 20~time steps, corresponding to $t_{\text{MD}} = 0.2 \, \sqrt{m \sigma^2/\varepsilon}$. Then, each trial configuration is either accepted or rejected using the Metropolis criterion applied to the bias potential only~\cite{Gelb2003}. We also use replica exchange moves~\cite{GeyerThompsonJAmAstat1995} between neighboring windows to improve convergence. After each evaluation of the bias potential, two neighboring windows are selected at random and an exchange is attempted between these two replicas.

\section{Results}
\label{sec:results_gem4}

In order to quantify nucleation and crystal growth timescales and select a regime were these timescales are accessible to numerical simulation, we start by determining the mobility of individual particles. The mobility is measured both in terms of the particles' mean-square displacement (MSD) as a function of time as well as the survival probability for two-particle clusters and its time derivative. Then, we calculate free energy curves as a function of the largest crystalline nucleus in the system for a range of pressures. Finally, we take a closer look at freezing trajectories taking the system from an undercooled liquid to a completely frozen state. This freezing takes place in two steps: the initial growth of a crystalline nucleus, followed by a solid-to-solid transition to the thermodynamically stable phase. We finally perform a committor analysis to investigate which conditions lead to the transformation of intermediate configurations to fully crystalline states.

\begin{figure}
\includegraphics[width=1.0\columnwidth]{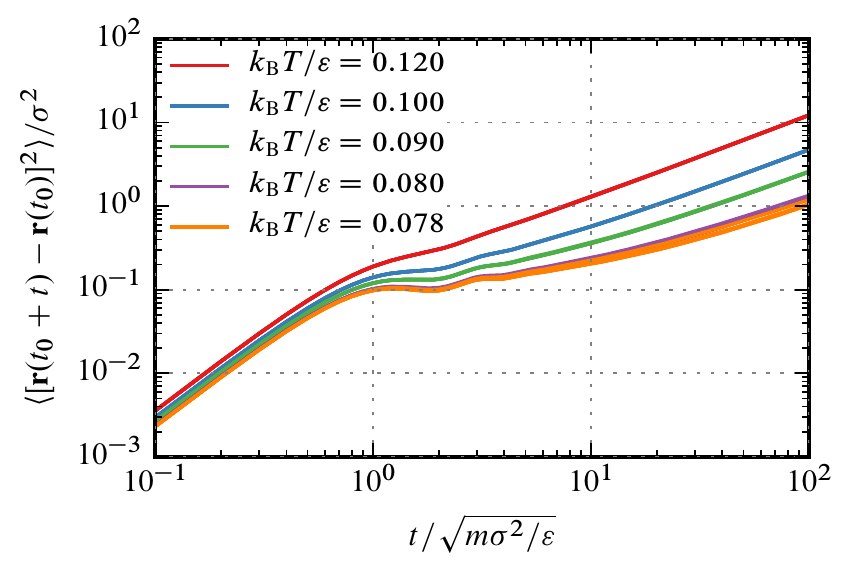}
\caption{Mean-square particle displacement in the liquid at a fixed density of $\rho \sigma^3 = 1$ as a function of temperature. For each temperature, a number of lines, each with a different reference time $t_0 / \sqrt{m \sigma^2 / \varepsilon} = \{100, 300, 700, 1500, 3100, 6300\}$, is shown. For low temperatures $k_{\text{B}} T / \varepsilon  \lesssim 0.08$, the MSD function takes a long time to converge, that is, the results have a slight dependence on the time of the measurement. Note that for temperatures $k_{\text{B}} T / \varepsilon  \leq 0.08$, simulation runs where the liquid turned (partly) into a crystal have been removed from the average.}
\label{fig:aging}
\end{figure}

\subsection{MSD and particle mobility}
\label{sec:gem4_msd}

For our numerical studies to be feasible, it is important that the initial undercooled liquid does not show slow, glassy dynamics and is well equilibrated. We quantify the degree of equilibration by measuring the mobility of particles. As we will see, for temperatures $k_{\text{B}} T / \varepsilon  \lesssim 0.08$, the particle mobility slightly depends on the total simulation time, which indicates the onset of glassy dynamics. Also, nucleation and crystalline growth timescales are strongly affected by low particle mobilities and glassy dynamics~\cite{Ikeda2011}. In the following, we will therefore avoid going to any lower temperatures.

We start by setting up the system in a completely random arrangement of particles. In other words, the initial configuration corresponds to the high-temperature limit. Then, we evolve the system forwards in time with the Langevin \textit{NVT} integrator. Immediately, we see the formation of a cluster liquid, with an average cluster occupancy very close to 2. Mean square displacements obtained by averaging over many independent simulation runs using $N~=~4000$ particles are displayed in Fig.\,\ref{fig:aging}. The results are corrected for the drift in the system's center of mass, which occurs due to the stochastic nature of the Langevin integrator.

The mean-square displacements
\begin{equation}
\Delta r^2(t) = \langle [\mathbf{r}(t_0 + t) - \mathbf{r}(t_0)]^2 \rangle
\end{equation}
are calculated for a number of different temperatures and reference times $t_0$. Of course, in an equilibrated system, $\Delta r^2(t)$ should not depend on the choice of $t_0$. However, as one can see in Fig.\,\ref{fig:aging}, that is not strictly the case for lower temperatures. In other words, for low temperatures, the system takes a long time to equilibrate, i.\,e., the measured values still depend on the time of the measurement long after the initial start of the simulation in a high-temperature configuration. Also, the average particle mobility is greatly reduced as particles, or rather, clusters, are trapped inside small regions of the total simulation volume. Conversely, due to the low or even vanishing free energy barrier for nucleation at low temperatures, the system begins to freeze into the crystalline phase. However, as a consequence of the reduced particle mobility, the freezing and particle re-arrangement takes a prohibitively long amount of time.

To avoid such slow dynamics, we will not go below temperatures of $k_{\text{B}} T / \varepsilon = 0.08$. For this temperature and the pressure and density range investigated in this work, cluster-fcc2 is the stable phase.

\subsection{Cluster survival probability and survival rate function}

For the conditions investigated in this study, even in the liquid phase, the average occupancy is very close to a value of two for all clusters. Therefore, it is reasonable to further characterize the particle mobility by calculating the cluster survival probability for clusters of size two as a function of time for different temperatures. For all clusters of two particles present at time~$0$, $p_2(t)$ gives the fraction of these clusters still present at time~$t$. This survival probability can be written as a correlation function,
\begin{equation}
p_2(t) = \frac{\langle h_2(0) h_2(t) \rangle}{\langle h_2 \rangle},
\end{equation}
where the indicator function $h_2(t)$ is $1$ if a particular pair of particles forms a 2-cluster and 0 otherwise, and $\langle h_2 \rangle$ is the equilibrium average $\int \mathrm{d}x \, h_2(x) e^{-\beta H(x)}$. Note that we only check if any particular cluster, consisting of two particles $i$ and $j$, still exists after a time $t$, regardless of what has happened in between. Also, a cluster is considered as not surviving if it has been joined by a third particle, even though this rarely happens in practice.

\begin{figure}
\includegraphics[width=1.0\columnwidth]{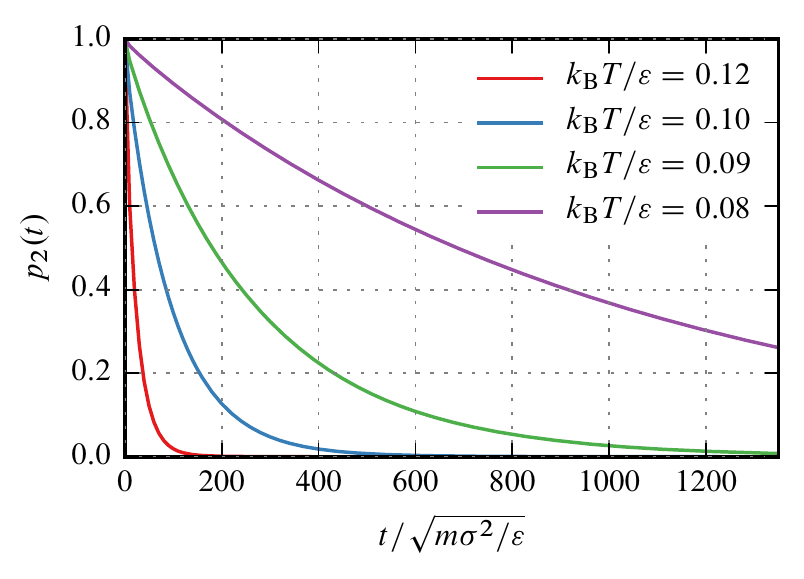}
\caption{Two-particle cluster survival function $p_2(t)$ in the liquid at a fixed density of $\rho \sigma^3 = 1$ as a function of temperature.}
\label{fig:survivalrate}
\end{figure}

In Fig.\,\ref{fig:survivalrate}, we have plotted the survival probability for clusters of two particles in the liquid at different temperatures. In the liquid, this function shows an exponential decay. However, in accordance with the results obtained from the particle MSD calculations in Sec.\,\ref{sec:gem4_msd}, the typical timescale for the decay strongly grows for lower temperatures. This illustrates why nucleation and crystalline growth timescales increase so much at low temperatures: not only are particles trapped inside cages formed by neighboring particles, they also do not leave their clusters any more. As a consequence, clusters have to move as a whole instead of single particles in order for re-arrangement to take place.

\begin{figure}
\includegraphics[width=1.0\columnwidth]{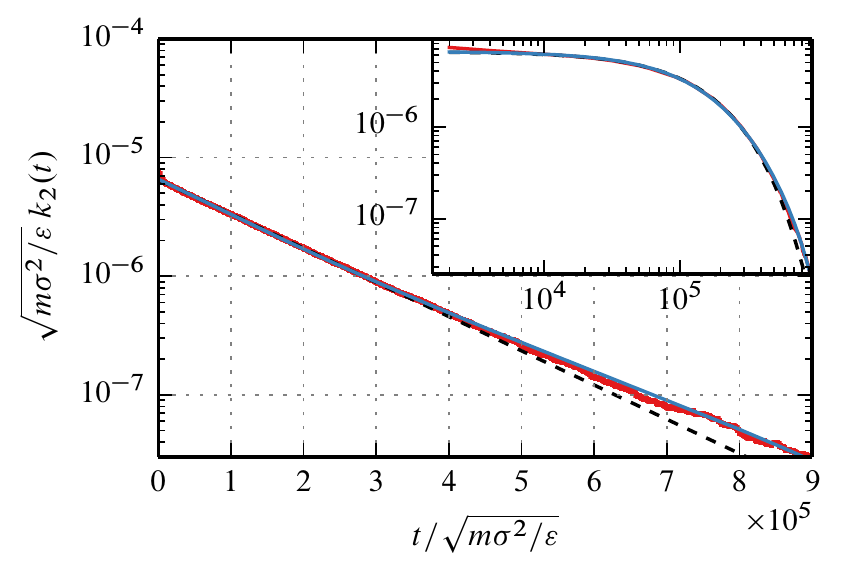}
\caption{Two-particle cluster survival rate function $k_2(t)$ in the fcc2 crystal at a pressure of $p \sigma^3 / \varepsilon = 1.49$ and a temperature of $k_{\text{B}} T / \varepsilon~=~0.08$. The blue line is a fit according to the inverse Laplace transform of Eq.\,\ref{eq:kofs}, and the dashed line is a fit with an exponential decay. Note the deviation from a purely exponential behavior for long timescales. Inset: same data plotted on a log--log scale.} 
\label{fig:koft}
\end{figure}

Not surprisingly, the timescales for cluster survival are much larger for crystalline structures. In the case of a cluster crystal consisting of 2-clusters, a particle can spontaneously hop to an adjacent lattice site, leading to a pair of a single particle and a 3-cluster. Subsequently, by chance, the same particle may hop back to its original lattice site, reforming the original 2-cluster. However, it is also possible that another particles takes its place, while the particle that has originally hopped eventually moves further away from its initial position in subsequent hopping events. Such a process can be modeled with a reaction-diffusion approach, initially developed to model hydrogen-bond kinetics~\cite{Luzar1996}. In particular, one aims to obtain an expression for the rate function
\begin{equation}
k_2(t) = - \frac{\mathrm{d} p_2(t)}{\mathrm{d} t},
\end{equation}
which is assumed to follow the form
\begin{equation}
\frac{\mathrm{d} p_2(t)}{\mathrm{d} t} = -k p_2(t) + k' n(t).
\end{equation}
Here, $n(t)$ is the probability that a particle initially part of a 2-cluster is still present at an adjacent lattice site, and $k$ and $k'$ are the rate constants for break-up and re-formation of 2-clusters, respectively. Then, it is assumed that the particle density $\rho(\mathbf{r}, t)$ changes by normal diffusion as well as the break-up and formation of clusters,
\begin{equation}
\frac{\partial}{\partial t} \rho(\mathbf{r}, t) = D \Delta \rho(\mathbf{r}, t) + \delta(\mathbf{r})[k p_2(t) - k' n(t)],
\end{equation}
where $D$ is the single-particle diffusion constant of the system. The model can be solved analytically in the Laplace domain. Thus, the rate function $k_2(t)$ is the inverse Laplace transform of~\cite{Luzar1996}
\begin{equation}
k_2(s) = \frac{k}{s + k + k' f(s)},
\label{eq:kofs}
\end{equation}
where
\begin{equation}
f(s) = 3 \tau \left[ 1 - \sqrt{s \tau} \arctan(1 / \sqrt{s \tau}) \right],
\end{equation}
and $\tau \propto 1 / D$ is a time constant related to particle diffusion in the system. We have calculated $p_2(t)$ for a perfect fcc2 crystal (Fig.\,\ref{fig:koft}), and fitted the data according to Eq.\,\ref{eq:kofs}. The coefficients we have obtained are $\tau = 8.7 \times 10^7 \sqrt{m \sigma^2/\varepsilon}$, $k = 6.6 \times 10^{-11} \sqrt{\varepsilon / m \sigma^2}$, and $k' = 6.7 \times 10^{-6} \sqrt{\varepsilon / m \sigma^2}$. As expected, $k \ll k'$, indicating that it is much easier to re-form a cluster than to break it up initially. The inverse Laplace transform is evaluated numerically~\cite{Stehfest1970}. As can be inferred from Fig.\,\ref{fig:koft}, the decay of $p_2(t)$ clearly deviates from exponential (and from power-law) behavior, but is reproduced well with the diffusion model. Note that one would expect an exponential decay, if there was zero probability for any broken-up cluster to re-form at a later time. However, as this may indeed happen, at long timescales the resulting rate function has a higher value than an exponential decay would predict.

\subsection{Nucleation free energies}
\label{sec:gem4_fe}

In a next step, using $NpT$ umbrella sampling, we have calculated equilibrium free energies for the nucleation from an undercooled liquid as a function of the size of the number of clusters in the largest crystalline nucleus $N_{\text{nuc}}$. The free energy is calculated from the probability density obtained via umbrella sampling, $F(N_{\text{nuc}}) = -k_{\text{B}} T \ln [\rho(N_{\text{nuc}})]$. The free energy calculations were carried out at constant pressure rather than at constant volume. This choice ensures that the system can better accommodate local density fluctuations that might be important in the liquid-to-solid nucleation. A number of independent simulations have been performed in order to improve sampling accuracy.

\begin{figure}
\includegraphics[width=1.0\columnwidth]{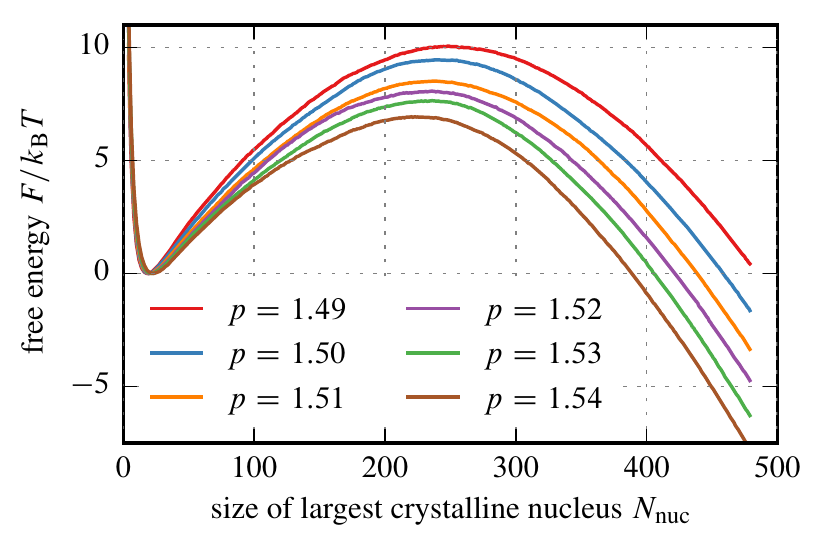}
\caption{Free energies for the nucleation from an undercooled liquid at $k_{\text{B}} T / \varepsilon = 0.08$ and different pressures ($p$ in units of $\varepsilon / \sigma^3$) as a function of the size $N_{\text{nuc}}$ of the largest crystalline nucleus. Pressures, from top to bottom, are $p \sigma^3 / \varepsilon = 1.49$ (red), 1.50 (light blue), 1.51 (orange), 1.52 (purple), 1.53 (green), and 1.54 (brown). All curves are normalized such that the local minimum for low nucleus size is located at $F / k_{\text{B}} T = 0$.}
\label{fig:fe_npt}
\end{figure}

Free energy curves as a function of nucleus size are shown in Fig.\,\ref{fig:fe_npt} for a temperature of $k_{\text{B}} T / \varepsilon = 0.08$ and a range of pressures. Clearly, for increasing the pressure, the size of the critical nucleus as well as the barrier height decrease. The local minimum to the left of the barrier present in all curves originates from the fact that the order parameter measures the size of the \textit{largest} crystalline nucleus. Hence, due to spontaneous fluctuations, even in the purely liquid phase, the most probable order parameter value is a number larger than zero. We use a value of $N_{\text{nuc}} = 480$ as upper limit in the free energy curves, which is well beyond the barrier for all conditions investigated. In particular, any simulation started with a nucleus of this size will spontaneously freeze into a fully crystalline state. The particle number $N = 16\,000$ used in the free energy calculations was chosen such that the critical nuclei fit well within the simulation box without coming too close to their own periodic images.

As a further step, one might ask whether the form of the free energy curves for nucleation from the liquid agrees with predictions from classical nucleation theory (CNT). However, in order to address this question, we first have to bring the free energy curves into a form suitable for comparison with CNT. Let us start by defining the function $\tilde{n}(n, N)$ as the (instantaneous) number of nuclei of size $n$ in a system of total size $N$. In order to compare with CNT, we need the quantity
\begin{equation}
f(n) = \langle \tilde{n}(n, N) \rangle,
\label{eq:f_of_n}
\end{equation}
that is, the average number of nuclei of size $n$ in a system of $N$ particles. $f(n)$ can also be interpreted as the (unnormalized) per-volume probability to find a nucleus of size $n$ in the system. In contrast, our free energies from the umbrella sampling simulations are computed from $\rho(n)$, the probability that the \textit{largest} crystalline nucleus in the system is of size $n$. The two quantities are only the same for large enough $n$, such that there is only a single nucleus of that size in the system. In Appendix\,\ref{ap:cnt_comparison}, we show how to obtain $f(n)$ in a systematic way. Effectively, one performs an additional, unconstrained simulation in the liquid state, and thus calculates $f(n)$ for small values of $n$. Then, this $f(n)$ is patched together with $\rho(n)$ at a carefully selected value of $n$.

Having obtained $f(n)$ and the respective free energy $G(n) = -k_{\text{B}} T \ln f(n)$, we can now compare their functional form to the prediction of CNT. As both the surface tension and the bulk chemical potential of our system are unknowns, we simply use a fitting function of the form
\begin{equation}
g(n) = a n^{2/3} + b n + c,
\end{equation}
where $a$ is the surface term, $b$ is the bulk contribution, and $c$ is just a normalization constant. Similarly, when introducing a size-dependent surface tension via the Tolman correction~\cite{Troester2012}, as well as a logarithmic correction for fluctuating nucleus shapes~\cite{Prestipino2012}, we use
\begin{equation}
g(n) = a n^{1/3} + b n^{2/3} + c n + d \ln n + e.
\end{equation}
Results for two selected conditions (the same as the ones in Fig.\,\ref{fig:greatest_nucleus_probility} in Appendix\,\ref{ap:cnt_comparison}) are shown in Fig.\,\ref{fig:patched_fe}. While the overall quality of both fits is satisfactory, the inclusion of the Tolman correction and the correction for fluctuations in the nucleus shape significantly improve the agreement with the free energy curves obtained numerically. Not surprisingly, neither method is able to reproduce the numerical results at very low nucleus sizes. There, the assumption of spherical nuclei---even if just in an average sense as it is assumed when applying the shape correction term---is certainly not justified.

\begin{figure}
\includegraphics[width=1.0\columnwidth]{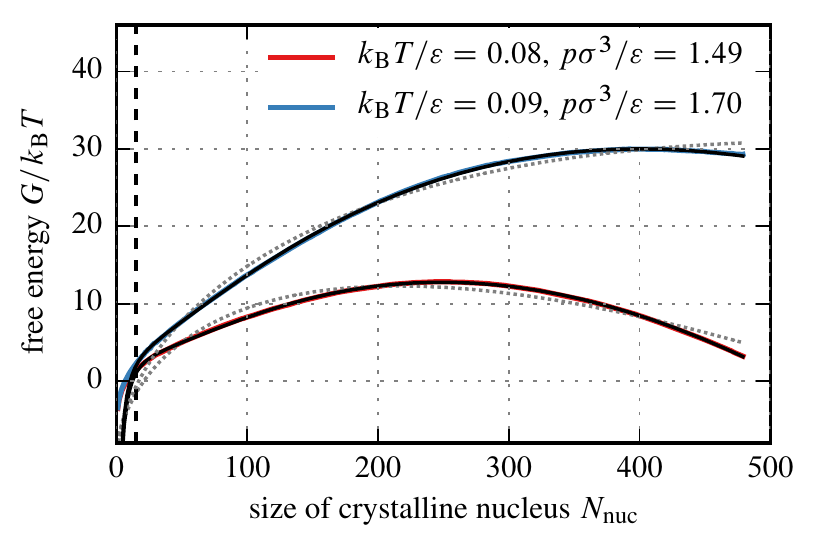}
\caption{Free energy curves $G(n)$ for two different conditions (red, blue), and numerical fits with the CNT prediction with (black) and without (grey, dotted) Tolman correction and nucleus shape correction. In both cases, the fit has been performed for values of $N_{\text{nuc}} \ge 15$ only, as indicated by the vertical dashed line.}
\label{fig:patched_fe}
\end{figure}

\subsection{Composition of nuclei: equilibrium}

\begin{figure}
\includegraphics[width=1.0\columnwidth]{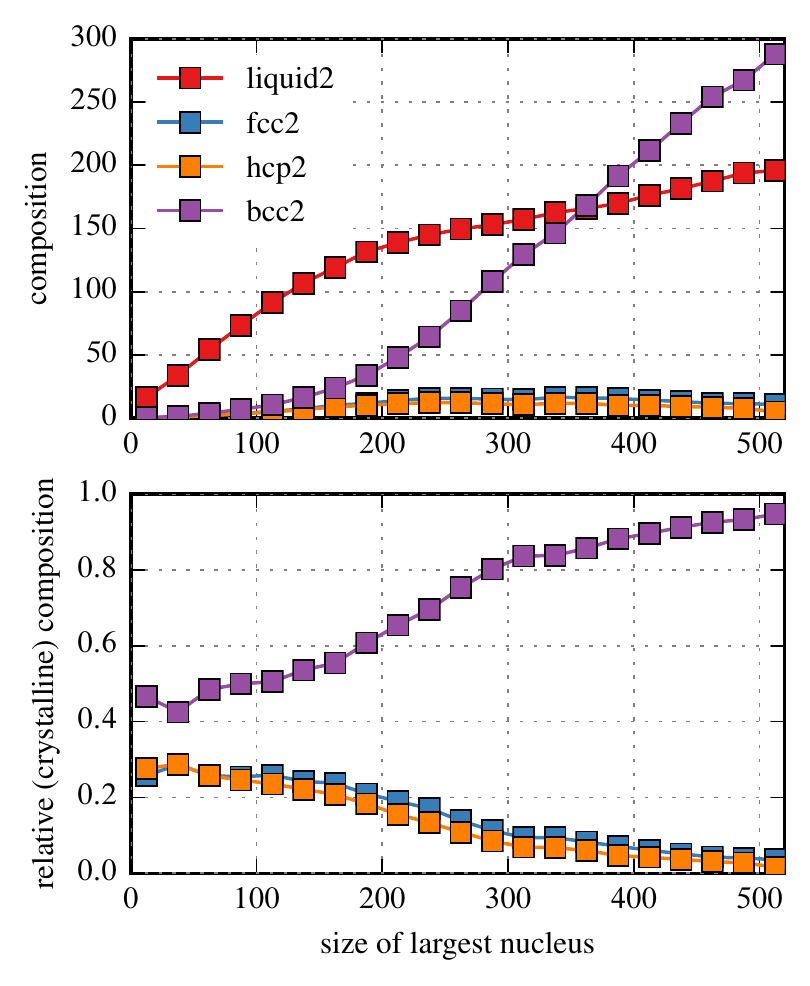}
\caption{(Top) Equilibrium nucleus composition at $k_{\text{B}} T / \varepsilon = 0.08$ and $p \sigma^3 / \varepsilon = 1.50$ (fcc2 stable) as a function of the nucleus size, measured in number of clusters. (Bottom) Relative fraction of the three crystalline structures fcc2, hcp2, and bcc2. In both panels, a significant change occurs around a nucleus size of 200 clusters.}
\label{fig:composition_fcc2}
\end{figure}

As shown in Sec.\,\ref{sec:gem4_fe}, the crystallization of the supercooled liquid is opposed by a free energy barrier and occurs via the formation of a small crystalline nucleus. However, this nucleus does not necessarily have the structure of the thermodynamically most stable one. Indeed, Ostwald's “step rule”~\cite{Ostwald1897} states that nucleation should first occur into the structure with the lowest free energy difference to the metastable liquid. For instance, a recent study by Mithen and coworkers~\cite{Mithen2015} has shown that in the Gaussian core model (GCM or GEM-2), the bcc structure is always favored in the early crystallization stages even at conditions at which the fcc-crystal is the stable phase. We observe a similar scenario in the GEM-4 system. In Fig.\,\ref{fig:composition_fcc2}, we have plotted the average nucleus composition as a function of nucleus size, for nuclei taken from our umbrella sampling simulations. The composition $c_i(N_{\text{nuc}})$ is defined as the average number of clusters of structure $i$ in a nucleus of size $N_{\text{nuc}}$, where $i$ is either one of liquid2, fcc2, hcp2, or bcc2. Since the bias applied during umbrella sampling only depends on the value of $N_{\text{nuc}}$, it does not affect the nucleus composition compared to an equilibrium simulation. For the conditions $k_{\text{B}} T / \varepsilon = 0.08$ and $p \sigma^3 / \varepsilon = 1.50$, which are used to calculate the compositions shown in Fig.\,\ref{fig:composition_fcc2}, cluster-fcc2 is the thermodynamically stable crystal structure. Nevertheless, the fraction of actual fcc2 clusters in small crystalline nuclei is very low and also approximately constant over a large range of nucleus sizes. Very small nuclei consist of about the same number of bcc2, and fcc2 and hcp2 clusters, respectively. However, a significant change in composition occurs around a nucleus size of 200: the relative fraction of fcc2 and hcp2 drops, while the fraction of bcc2 increases by the same amount. At the same nucleus size, we see a change of the slope for the curve representing liquid clusters. These clusters are mainly found on the surface of the nuclei. A typical example for such a nucleus, consisting primarily of liquid-like clusters on the surface, as well as bcc2, fcc2, and hcp2 clusters in the core, is shown in Fig.\,\ref{fig:bnucleus}.

\begin{figure}
\includegraphics[width=1.0\columnwidth]{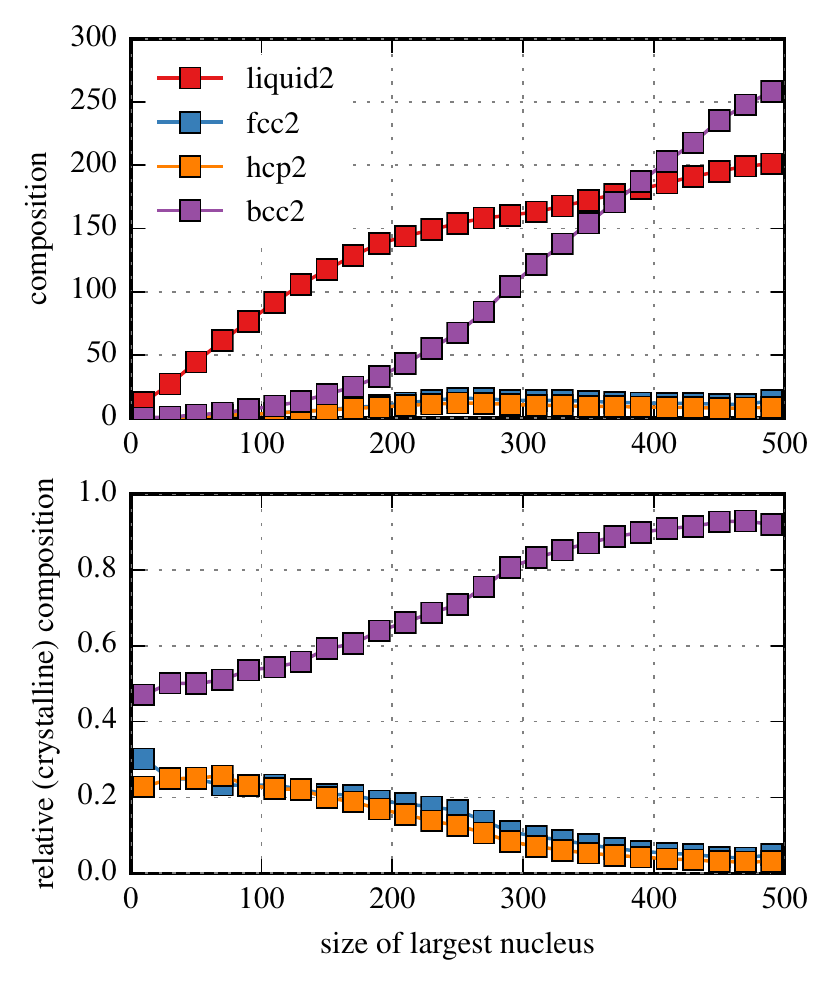}
\caption{(Top) Equilibrium nucleus composition at $k_{\text{B}} T / \varepsilon = 0.09$ and $p \sigma^3 / \varepsilon = 1.70$ (bcc2 stable) as a function of the nucleus size, measured in number of clusters. (Bottom) Relative fraction of the three crystalline structures fcc2, hcp2, and bcc2.}
\label{fig:composition_bcc2}
\end{figure}

By increasing both temperature and pressure to $k_{\text{B}} T / \varepsilon~=~0.09$ and $p \sigma^3 / \varepsilon~=~1.70$, one can reach a regime where bcc2 is actually the thermodynamically favored phase. However, as we see in Fig.\,\ref{fig:composition_bcc2}, the average nucleus composition is practically identical to the case where fcc2 is stable. This is another hint that the strong preference for the bcc2 structures in the GEM-4 system is a kinetic effect.

\begin{figure}
\includegraphics[width=0.7\columnwidth]{{{bnucleus.68}}}
\caption{Critical nucleus at $k_{\text{B}} T / \varepsilon = 0.08$ and $p \sigma^3 / \varepsilon = 1.50$ (surrounding liquid particles not part of the nucleus are not shown). Particles are colored according to their local structure in red (liquid2), light blue (fcc2), orange (hcp2), or purple (bcc2). The outlines of the Voronoi cells used in the order parameter calculation are drawn as thin blue lines. Even though fcc2 is the thermodynamically stable structure for these conditions, the nucleus consists of more bcc2 clusters than fcc2 clusters (see Fig.\,\ref{fig:composition_fcc2}).}
\label{fig:bnucleus}
\end{figure}

\subsection{Composition of nuclei: freezing trajectories}
\label{sec:gem4_freezing}

\begin{figure}
\includegraphics[width=1.0\columnwidth]{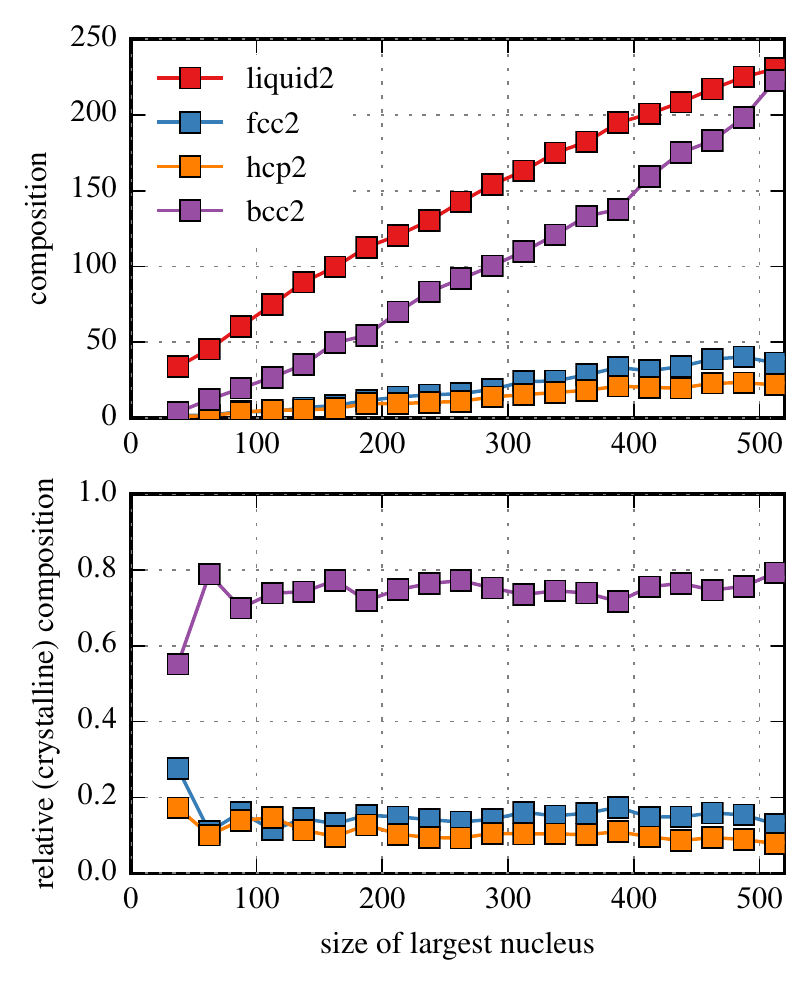}
\caption{(Top) Nucleus composition evaluated for freezing trajectories at $k_{\text{B}} T / \varepsilon = 0.08$ and $p \sigma^3 / \varepsilon = 1.50$ as a function of the nucleus size, measured in number of clusters. (Bottom) relative fraction of the three crystalline structures fcc2, hcp2, and bcc2.}
\label{fig:composition_freezing_fcc2}
\end{figure}

\begin{figure}
\includegraphics[width=1.0\columnwidth]{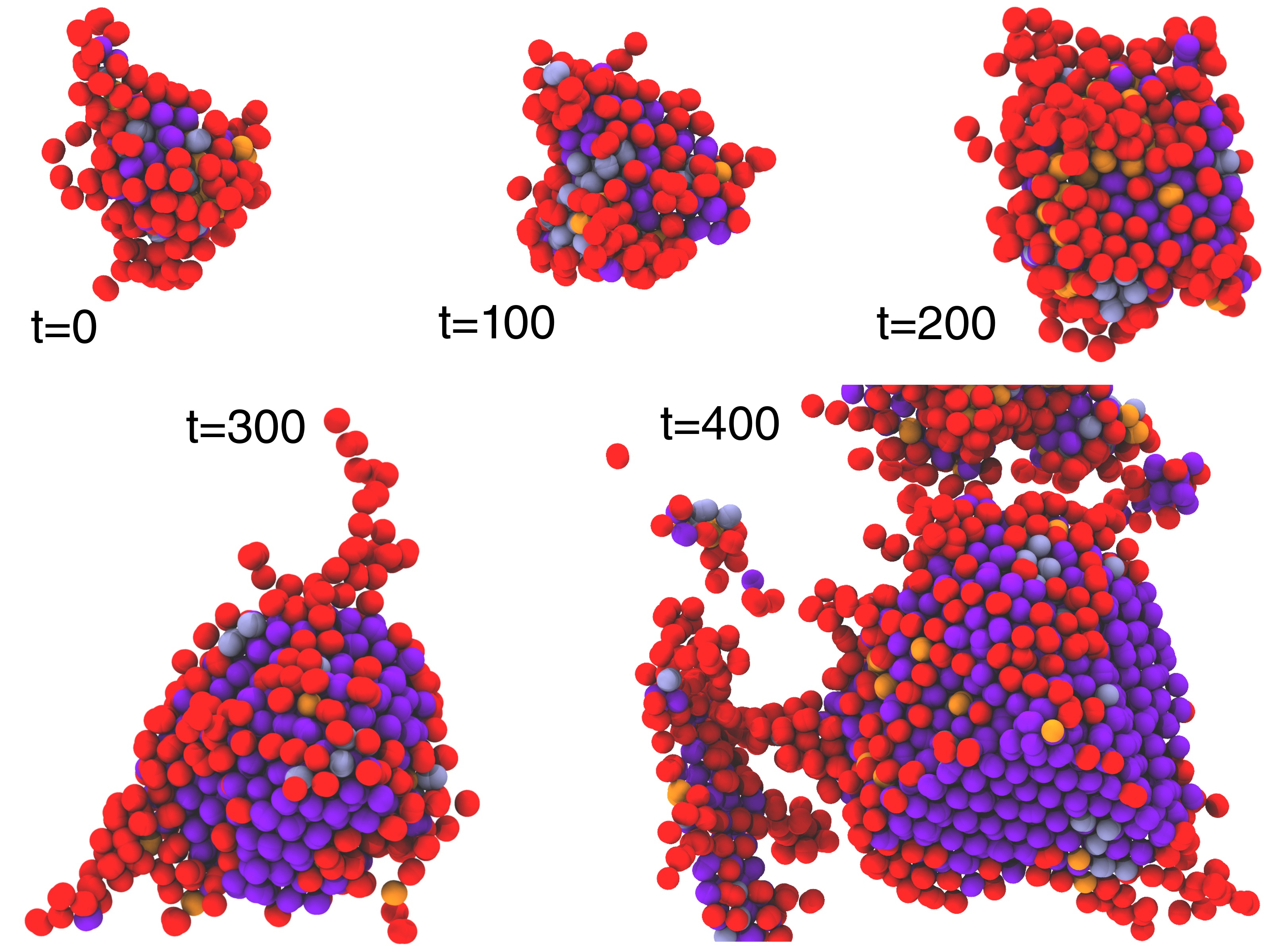}
\caption{Snapshots of the largest crystalline nucleus along a freezing trajectory. Times are given in units of $\sqrt{m \sigma^2/\varepsilon}$. At $t=400$, the nucleus is large enough to interact with its own periodic images.}
\label{fig:sequence}
\end{figure}

In order to investigate if there is any difference in nucleus composition between equilibrium configurations and those taken from complete freezing pathways, we have started a number of trajectories from configurations taken from the top of the free energy barrier (see Fig.\,\ref{fig:fe_npt}). While in some of these trajectories the nucleus disappears, in others it grows and eventually the entire system crystallizes. For these freezing trajectories, especially for small nuclei, the relative fraction of bcc2 is significantly larger than in the nuclei drawn from the equilibrium distribution (Fig.\,\ref{fig:composition_freezing_fcc2}). In other words, the presence of a bcc2 nucleus strongly increases the probability of a configuration to freeze, in accordance with direct observations of freezing trajectories. In particular, small fcc2 structures present in early nuclei often transform into bcc2. Then, more often than not, the whole system freezes into a state consisting primarily of bcc2 particles. A typical bcc2-rich freezing trajectory is shown in Fig.\,\ref{fig:sequence}. The transformation to the thermodynamically favored fcc2 phase happens only after that, on a much longer timescale than the growth of the initial nucleus. We will look into these later stages of the transformation in more detail in the following.

\subsection{Full freezing trajectories and committor analysis}

\begin{figure}
\includegraphics[width=1.0\columnwidth]{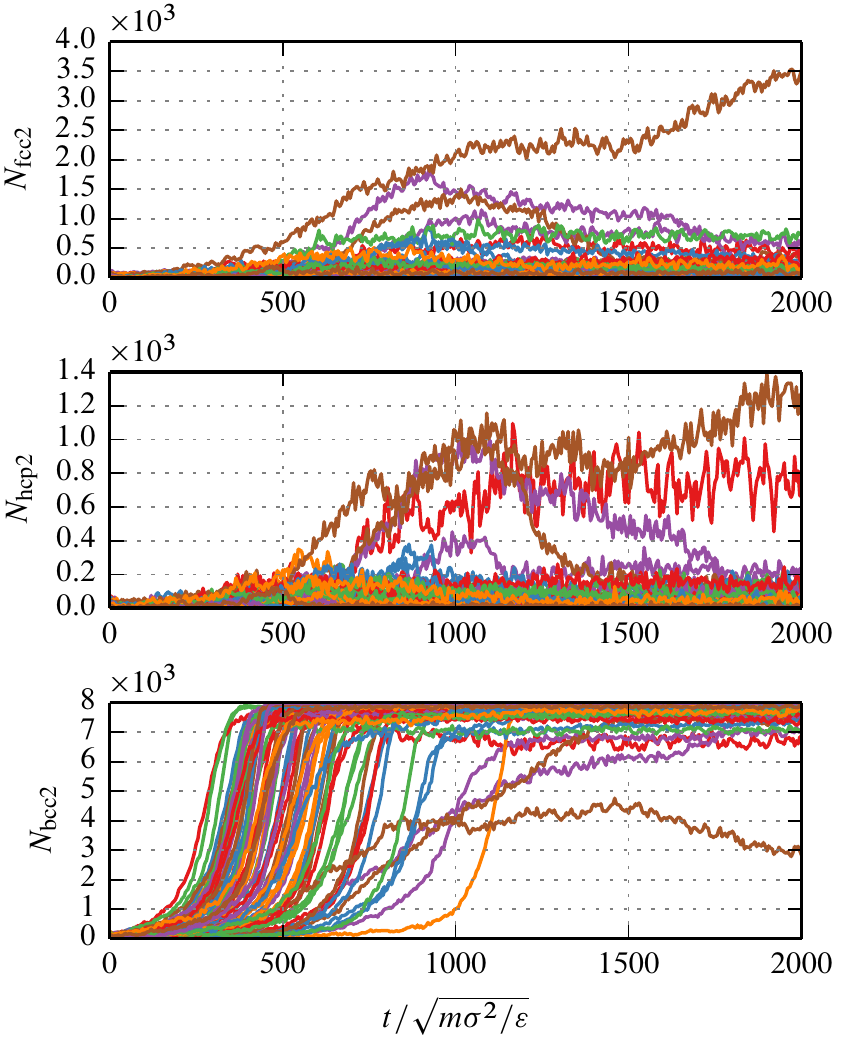}
\caption{Number of fcc2, hcp2, and bcc2 clusters in the largest crystalline nucleus for a selection of freezing trajectories starting from a nucleus size of 250 clusters, at the top of the free energy barrier. The simulations were run at $k_{\text{B}} T / \varepsilon = 0.08$, $p \sigma^3 / \varepsilon = 1.49$, and a system size of $N = 16\,000$ particles.}
\label{fig:freezing}
\end{figure}

\begin{figure}
\includegraphics[width=1.0\columnwidth]{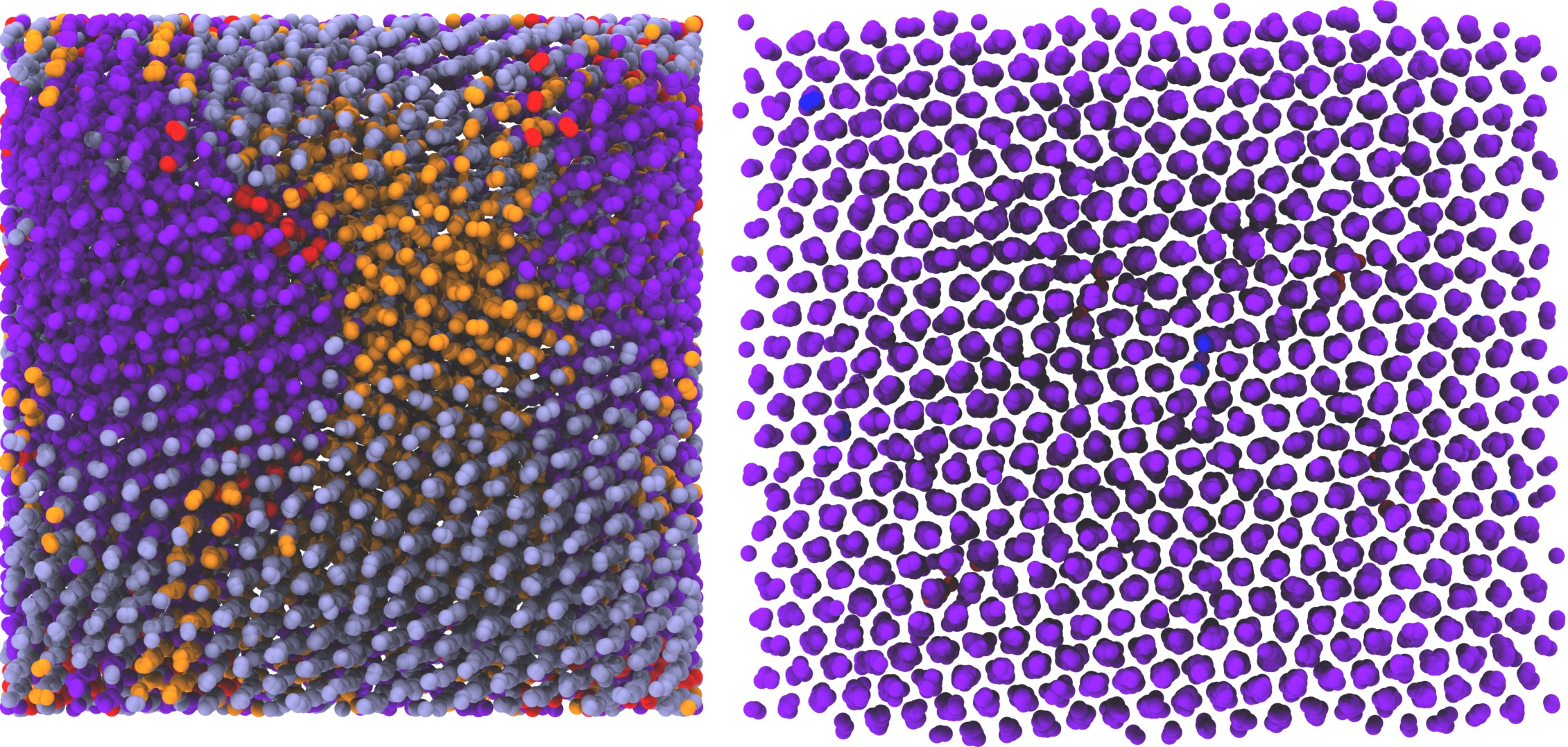}
\caption{Cross section through the simulation box at the end of two freezing trajectories. The trajectories ended in a polymorph with a high number of fcc2 clusters (left) and in a pure bcc2 structure (right). Particles are colored red (liquid2), light blue (fcc2), orange (hcp2), and purple (bcc2).}
\label{fig:cuts_polymorphs}
\end{figure}

As a next step in understanding the mechanisms of crystal growth in the GEM-4 system, we take a closer look at  freezing trajectories. So far, we have already seen that both for equilibrium configurations on the free energy barrier as well as as for configurations taken from trajectories that are ending in a fully crystalline system, the bcc2 structure is strongly favored for small crystalline nuclei. Hence, the question arises whether this trend continues for larger nuclei as well, up to the complete crystallization of the entire system. In principle, in situations where fcc2 is the thermodynamically stable structure, two distinct scenarios for the crystal growth process are possible: in one case, the structure turns into fcc2 in the interior of the crystalline nucleus already while the system is still partially in a liquid state. Such behavior was observed, for example, in a Lennard-Jones mixture~\cite{Jungblut2011}. In the other scenario, the system first completely freezes to bcc2, and only later, in a second step, transforms to fcc2. As already hinted at in Sec.\,\ref{sec:gem4_freezing}, our results show that for GEM-4 under the conditions investigated, the two-step freezing scenario is the dominating one. In Fig.\,\ref{fig:freezing}, we plot the instantaneous composition of the largest crystalline nucleus along freezing trajectories started from the critical nucleus size. The large majority of trajectories ends up in a state where the entire system is in a bcc2 structure. Also, as indicated by the more or less parallel slopes during the bcc2 growth phase (bottom panel of Fig.\,\ref{fig:freezing}), the growth of the bcc2 structure happens with a constant rate, which is much faster than the typical growth rate for fcc2 or hcp2. Hence, the system simply does not have enough time to transform to fcc2, and transforms completely to bcc2 instead. Often, the fcc2 region in the nucleus actually shrinks during the growth process. Among the examples shown in Fig.\,\ref{fig:freezing}, only a single trajectory (plotted in brown) has managed to overcome this bcc2 dominance early on in the growth process. However, even in this trajectory, a significant amount of bcc2 as well as hcp2 clusters remain in the system at the end of the simulation.  We have illustrated this final state in Fig.\,\ref{fig:cuts_polymorphs} (left), along with a more typical bcc2-only final state (right). The configuration containing many fcc2 clusters is very similar to the typical configurations observed by Mithen and coworkers~\cite{Mithen2015} for the freezing in the Gaussian core model, which are also characterized by a high degree of polymorphism. In contrast, the majority of bcc2-only final configurations observed in our simulations are rather uniform.

When simulating for a longer time, the fcc2-rich trajectory finally transforms into a state where there is no significant amount of bcc2 structure left. However, as the resulting fcc2 crystal is not aligned with the simulation box, a number of stacking faults remain. Apparently, the survival of a large enough fcc2 structure during the initial crystalline growth phase is crucial for the full transformation of the entire system. Eventually, also a bcc2 crystal will transform into the stable fcc2 form, but the timescales required for this transformation are larger than the time accessible to our simulations. Note that it is not possible to set up a finite simulation volume with periodic boundary conditions and a fixed number of particles that allows the formation of both a perfect fcc2 as well as a perfect bcc2 crystal. Therefore, as a compromise, we have used a particle number which is a perfect fit neither for fcc2 nor for bcc2. 

\begin{figure}
\includegraphics[width=1.0\columnwidth]{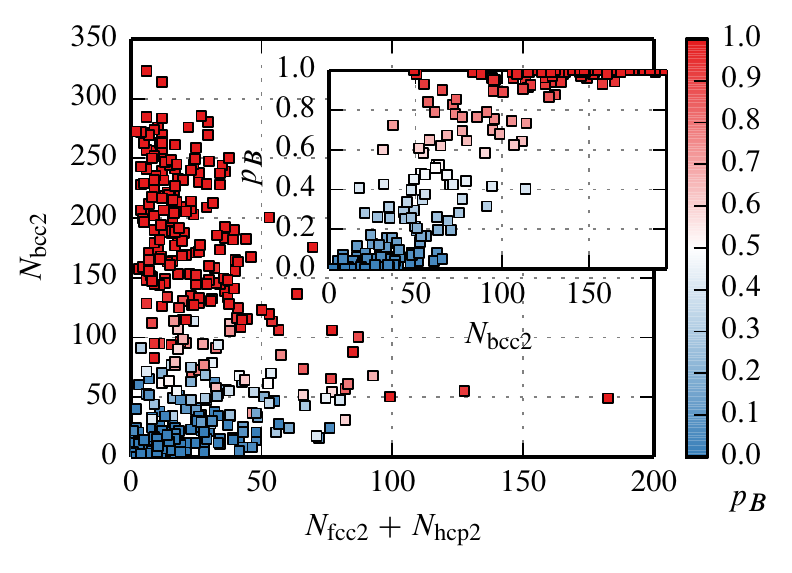}
\caption{(Color coded) Committor as a function of the composition of the largest crystalline nucleus. Inset: committor plotted as a function of $N_{\text{bcc2}}$ in the largest crystalline nucleus (same data set). The same conditions as for the trajectories shown in Fig.\,\ref{fig:freezing} were used. The algorithm described in Ref.\,\onlinecite{tps_fluid} was employed for the calculation, using ${N_{\text{min}} = 100}$ and ${N_{\text{max}} = 500}$.}
\label{fig:committors}
\end{figure}

So far, we have only looked at the properties of freezing trajectories, i.\,e., trajectories that end up in a crystalline state after a rather short simulation time. However, the question remains whether we can predict if any given intermediate crystallite will freeze or will instead melt again. Quantitatively, a committor analysis is able to provide a measure of freezing affinity for any given configuration. In a system with two (meta-) stable states $A$ and $B$, the committor $p_B(\mathbf{x})$ of configuration $\mathbf{x}$ is the fraction of dynamical pathways started from $\mathbf{x}$ that first reach state $B$~\cite{tps0, tps_fluid}. In accordance with our free energy calculation, for our system we define the region $B$ in terms of the number of clusters in the largest crystalline nucleus. The system is considered to be in state $B$ if $N_{\text{nuc}} > 500$, while we define state $A$ by $N_{\text{nuc}} < 25$. Hence, the committor is the probability of a configuration to freeze. In Fig.\,\ref{fig:committors}, we show the committor plotted as a function of the composition of the largest crystalline nucleus. Clearly, there is a strong relation between the committor value and the number of bcc2 clusters in the nucleus. This observation is further reinforced by the inset of Fig.\,\ref{fig:committors}, where we plot the committor against the number of bcc2 clusters in the largest crystalline nucleus. In this representation, the committor shows a sigmoidal shape, indicating that the number of bcc2 clusters is a useful measure for the progress of the freezing transition. Note however that due to the exclusion of all but the bcc2 clusters, the midpoint is shifted to lower values compared to the top of the free energy barrier shown in Fig.\,\ref{fig:patched_fe}. Furthermore, there are some configurations which have a high committor despite consisting of a rather low number of bcc2 clusters. These are the configurations which have more fcc2 and hcp2 clusters instead. Thus, even though such nuclei are occurring rarely, the presence of a large number of clusters in an fcc2 or hcp2 environment is favorable for further crystallization as well. However, some configurations exist which have a low committor despite consisting of many crystalline clusters, indicating that the nucleus size alone is not sufficient for a full description of the nucleation mechanism. In the following, we will show that the overall shape of the nuclei plays a role in determining the committor as well.

\begin{figure}
\includegraphics[width=1.0\columnwidth]{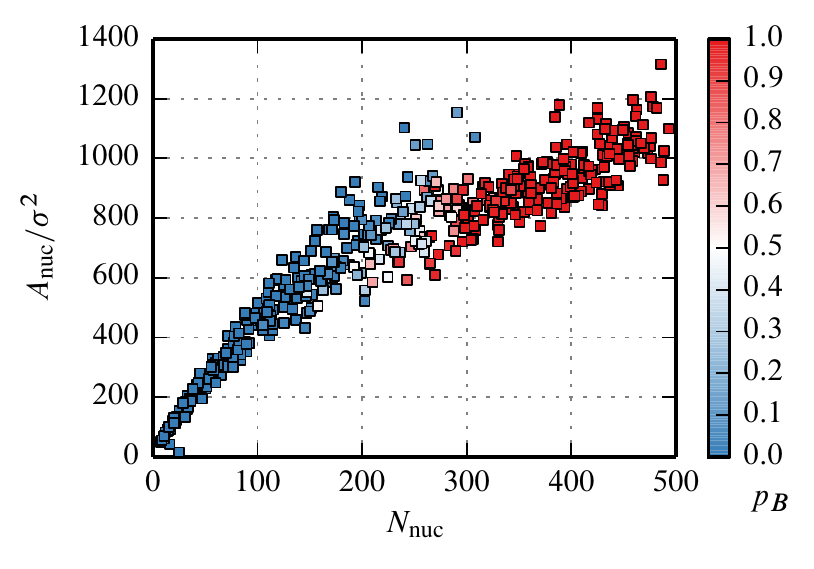}
\caption{(Color coded) Committor as a function of the total number of clusters and the surface area of the largest crystalline nucleus. The same data set as in Fig.\,\ref{fig:committors} is shown.}
\label{fig:surface_committors}
\end{figure}

We have plotted the same data set as a function of the total nucleus size including “liquid” surface particles and the surface area in Fig.\,\ref{fig:surface_committors}. The surface area of the nucleus, $A_{\text{nuc}}$, is defined as the sum over all Voronoi facet areas shared with a neighboring cluster that is not part of the nucleus. Note that in the vicinity of the critical nucleus size, around a value of $N_{\text{nuc}} = 250$, practically any committor value occurs. However, configurations with low committors tend to have a large surface area, while configurations with a low surface area have a higher committor. This behavior can be understood in terms of the shape of the nuclei: while compact, more or less spherical nuclei have a rather small surface area, the nuclei with larger surface area are much more open and fractal-like. Hence, for at a given total nucleus size, a compact structure is favorable for further crystallization.

\section{Discussion}
\label{sec:discussion_gem4}

In this work, we have investigated the formation and growth of cluster crystal phases from an undercooled liquid in the GEM-4 model. To distinguish between different ordered and unordered structures, we have employed bond-order parameters based on a Voronoi tesselation. This method has the advantage that it is (almost) free of arbitrary parameters and is still able to separate different crystal structures with high precision. Using molecular dynamics in the \textit{NpT} ensemble combined with umbrella sampling, we have studied crystallization over a range of pressures and two different temperatures. We have compared conditions where either the fcc2 or bcc2 structure, each with two particles per lattice site, is the thermodynamically stable one. Also in the undercooled liquid at the same conditions, typically two particles sit on top of each other. Due to caging effects, the particles' mobility in the liquid phase is strongly temperature-dependent. Hence, for cluster formation studies to be computationally feasible, the temperature has to be selected with great care to ensure a sufficiently high mobility. The mobility of particles is reduced even more in the crystalline phase, but less so than in a regular crystal because the main diffusion mechanism is the hopping of particles to adjacent lattice sites, rather than the movement of lattice defects. The kinetics of the particle hops can be reproduced well with a simple model that combines spontaneous break-up of clusters with diffusion.

In the crystallization of the GEM-4 system there is a strong preference for the formation of the bcc2 structure, even in cases where fcc2 is thermodynamically stable. The inspection of trajectories indicates that the presence of a bcc2-rich crystalline nucleus strongly increases the freezing probability of a given configuration. This kinetic preference translates to later stages of crystalline growth as well. More specifically, the growth rate for bcc2 is larger than for fcc2, such that most freezing configurations end up in a bcc2-only state, even if they initially contain a considerable amount of clusters in an fcc2-environment as well. Conversely, for the conditions investigated in our work, the final transformation to fcc2 is rarely observed. In future work we plan to investigate the mechanism of the bcc2-to-fcc2 solid-to-solid transition.

\begin{acknowledgments}
We thank Wolfgang Lechner and Bianca Mladek as well as Georg Menzl for useful discussions. The Initiativkolleg “Computational Science” of the University of Vienna and the Austrian Science Fund (FWF) within the SFB ViCoM (grant no. F41) are gratefully acknowledged for financial support. The computational results presented have been achieved using the Vienna Scientific Cluster (VSC).
\end{acknowledgments}

\appendix

\section{Center-of-mass calculation under periodic boundary conditions}
\label{ap:com_pbc}

The algorithm we use provides a straightforward and unambiguous way to calculate the center of mass of an arbitrary arrangement of particles in a system with periodic boundary conditions. We will outline the main result of Ref.\,\onlinecite{Bai2008} to calculate the center of mass of a system of $N$ particles of equal mass in a one-dimensional periodic system of size $x_{\text{max}}$. Naturally, for three-dimensional systems, the calculations have to be performed for each dimension separately.

We start by mapping each coordinate to an angle
\begin{equation}
\theta_i = 2 \pi \frac{x_i}{x_{\text{max}}}.
\end{equation}
Then, this angle is interpreted to be on a unit circle, and the corresponding two-dimensional coordinates are calculated,
\begin{align}
\xi_i &= \cos(\theta_i),\nonumber\\
\zeta_i &= \sin(\theta_i).
\end{align}
Now, we calculate the standard center of mass in the two-dimensional space,
\begin{align}
\overline{\xi} &= \frac{1}{N} \sum_{i=1}^{N} \xi_i,\nonumber\\
\overline{\zeta} &= \frac{1}{N} \sum_{i=1}^{N} \zeta_i.
\end{align}
Finally, we can map back the common center to an angle
\begin{equation}
\overline{\theta} = \mathrm{atan2}(-\overline{\zeta},-\overline{\xi}) + \pi,
\end{equation}
and then map back this average angle to a length coordinate in the range $[0, x_{\text{max}})$:
\begin{equation}
x_{\text{COM}} = x_{\text{max}} \frac{ \overline{\theta}}{2 \pi}.
\end{equation}
Here, $\mathrm{atan2}(y, x)$ is the two-argument inverse tangent function implemented in many programming languages and computer algebra systems. The negation of the arguments in combination with the shift of the function by $\pi$ ensures that $\overline{\theta}$ falls within $[0, 2 \pi)$.

We stress again that the algorithm described above is completely unambiguous, even for cases where the mass distribution is wide in comparison to the periodic box. This is not true when trying to calculate the center of mass for such a case using the usual minimum image convention with respect to some more or less random reference point. The algorithm will only fail in the case of a completely uniform mass distribution, for which the center of mass is not defined in a periodic system. Even then, the algorithm will return \emph{some} value (depending on the implementation of atan2), which is as good as any other for that particular situation.

\section{Comparison with classical nucleation theory}
\label{ap:cnt_comparison}

Here, we briefly recapitulate how one can compare results to classical nucleation theory (CNT) when using the size of the largest crystalline nucleus as the order parameter for free energy calculations.

Let us denote with $\rho(n)$ the probability that the largest nucleus in a system is of size $n$. The free energies calculated using umbrella sampling (Fig.\,\ref{fig:fe_npt}) are nothing but $F(n) = -k_{\text{B}} T \ln \rho(n)$. On the other hand, $f(n) = \langle \tilde{n}(n, N) \rangle$, is the average number of nuclei of size $n$. For comparing the two different functions, we can actually calculate $\rho(n)$ from $\langle \tilde{n}(n, N)\rangle$. Assuming that the nuclei are all independent of each other and hence their occurrence numbers in any single configuration are Poisson distributed around the known average, we have
\begin{equation}
\rho(n) = e^{- \sum_{i=n+1}^{\infty} \langle \tilde{n}(i, N) \rangle } [ 1 - e^{-\langle \tilde{n}(n, N) \rangle } ].
\label{eq:rho_of_n}
\end{equation}
In words, the probability that the largest nucleus in the system is of size $n$ is the probability that there is not a single nucleus of any larger size, times the probability that there is a least one nucleus of size $n$.

\begin{figure}
\includegraphics[width=1.0\columnwidth]{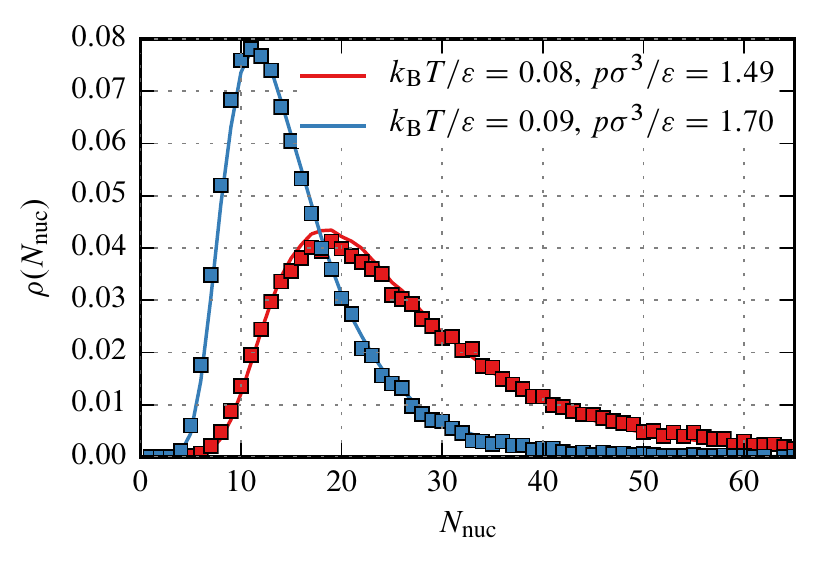}
\caption{$\rho(n)$, the probability to find a largest nucleus of size $n$, calculated for two different conditions and a system size of $N~=~16\,000$. Shown is a comparison between direct calculation with umbrella sampling (solid lines), and calculation via Eq.\,\eqref{eq:rho_of_n} (square symbols).}
\label{fig:greatest_nucleus_probility}
\end{figure}

\begin{figure}
\includegraphics[width=1.0\columnwidth]{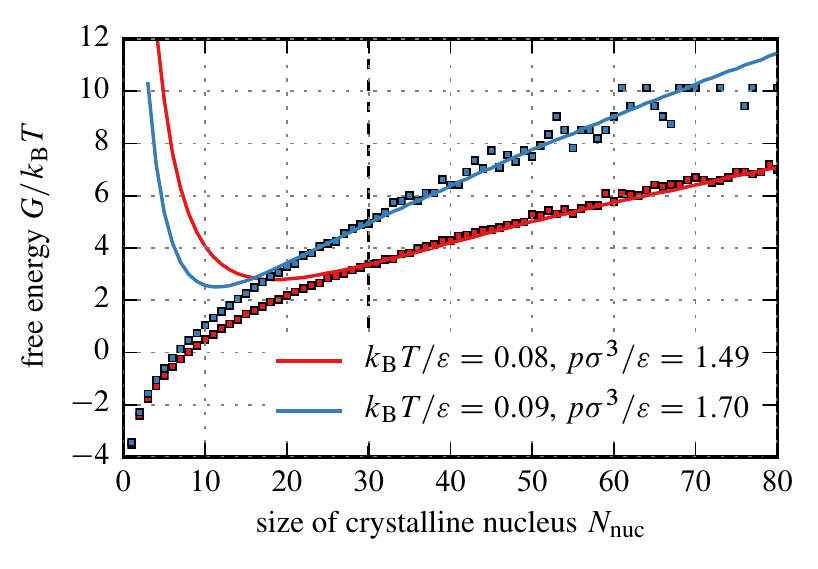}
\caption{Free energy from umbrella sampling (solid lines) and calculated by evaluating $-\ln \langle \tilde{n}(n, N) \rangle$ (symbols). The final result is patched together at a nucleus size of 30, indicated by a dashed line.}
\label{fig:patching}
\end{figure}

In Fig.\,\ref{fig:greatest_nucleus_probility}, we show a comparison between the two ways of calculating $\rho(n)$ for small values of $n$. To obtain these data, we only had to perform additional simulations of an unconstrained liquid system in order to sample $\langle \tilde{n}(n, N) \rangle$. Clearly, the calculation using Eq.\,\eqref{eq:rho_of_n} agrees well with the direct method using umbrella sampling, suggesting that the nuclei appear independently of each other. Note also that while the umbrella sampling results are normalized such that the total probability is 1, using Eq.\,\eqref{eq:rho_of_n} we directly obtain real probabilities for the occurrence of a largest nucleus of size $n$, without any further need for normalization. As a last step, we have to determine where to patch the two free energy curves together. We will use
\begin{equation}
G / k_{\text{B}} T = \begin{cases} -\ln f(n) &\quad n \leq n_{\text{patch}},\\
-\ln \rho(n) + k &\quad n > n_{\text{patch}},
\end{cases}
\end{equation}
where $ n_{\text{patch}}$ is the value where the two curves are stitched together and the constant $k$ is chosen such that the free energy is continuous at the stitching point. To find the point where to patch the two parts together, we plot both curves (Fig.\,\ref{fig:patching}) and select the stitching point visually. In conclusion, for comparison with CNT, we patch together $f(n)$ with $\rho(n)$ at a value of $ n_{\text{patch}}=30$, where the two curves are already practically indistinguishable from each other for all conditions investigated.

\end{document}